\documentclass[manuscript]{aastex}
\usepackage{spr-astr-addons}
\usepackage{url}
\usepackage{natbib}
\usepackage{amssymb, amsmath}
\def\kms{km~s$^{-1}$}

\usepackage{verbatim}
\usepackage{color}

\newcommand{\vsini}{\ensuremath{v_{{\mathrm e}}\sin i}}

\RequirePackage{color}
\begin{document}

\title{The nature of the late B-type stars HD 67044 and HD 42035}
\shorttitle{HD67044 and HD 42035}
\shortauthors{Monier et al.}

\author{R. Monier\altaffilmark{1}\altaffilmark{2}}
\affil{LESIA, UMR 8109, Observatoire de Paris, place J. Janssen, Meudon.}\affil{Laboratoire Lagange, Universit\'e de Nice Sophia, Parc Valrose,            06100 Nice, France.}
\and
\author{M. Gebran\altaffilmark{3}}
\affil{Department of Physics and Astronomy, Notre Dame University-Louaize, PO Box 72, Zouk Mikael, Lebanon.}
\and
\author{F. Royer\altaffilmark{4}}
\affil{GEPI, Observatoire de Paris, place J. Janssen, Meudon, France.}

\email{Richard.Monier@obspm.fr}


\begin{abstract}
While monitoring a sample of apparently slowly rotating superficially normal bright late B and early A stars in the northern hemisphere, we have discovered that HD 67044 and HD 42035, hitherto classified as normal late B-type stars, are actually respectively a new chemically peculiar star and a new spectroscopic binary containing a very slow rotator HD 42035 S with ultra-sharp lines  ($\vsini = 3.7$ km\,s$^{-1}$) and a fast rotator HD 42035 B with broad lines.
The lines of  \ion{Ti}{2}, \ion{Cr}{2}, \ion{Mn}{2},   \ion{Sr}{2} , \ion{Y}{2}, \ion{Zr}{2} and \ion{Ba}{2} are conspicuous features  in the high resolution SOPHIE spectrum ($R=75000$) of HD 67044. The \ion{Hg}{2} line at 3983.93 \AA\ is also present as a weak feature.
The composite spectrum of HD 42035 is characterised by very sharp lines formed in HD 42035 S superimposed onto the shallow and broad lines of HD 42035 B.
These very sharp lines
are mostly due to light elements from C to Ni, the only heavy species definitely present are  strontium and barium.
Selected lines of 21 chemical elements from He up to Hg  have been synthesized using model atmospheres computed with ATLAS9 and the spectrum synthesis code SYNSPEC48 including  hyperfine structure of various isotopes when relevant. These synthetic spectra have been  adjusted to high resolution high signal-to-noise spectra of HD 67044 and HD 42035 S in order to derive abundances of these  key elements.
HD 67044 is found to have distinct enhancements of  Ti, Cr, Mn,  Sr, Y, Zr, Ba and Hg and underabundances in He, C, O, Ca and Sc which shows that this star is not a superficially normal late B-type star, but actually is a new CP  star most likely of the HgMn type. HD 42035 S has provisional underabundances of the light elements from C to Ti and overabundances of heavier elements (except for Fe and Sr which are also underabundant) up to barium. These values are lower limits to the actual abundances as we cannot currently place properly the continuum of HD 42035 S. More accurate fundamental parameters and abundances for HD 42035 S and HD 42035 B will be derived if we manage to disentangle their spectra. They will help clarify the status of the two components in this interesting new spectroscopic binary.
\end{abstract}


\section{Introduction}
\label{sec:intro}
 We have recently undertaken a spectroscopic survey of all apparently slowly rotating bright early A stars (A0-A1V) and late B stars (B8-B9V) observable from the northern hemisphere. The incentive is 
to search for rapid rotators seen pole-on or new chemically peculiar B and A stars which have thus far remained unnoticed.
This project addresses fundamental questions of the physics of late-B and early-A stars: i) can we find new instances of rapid rotators seen pole-on (other than Vega) and study their physical properties (gradient of temperature across the disk, limb and gravity darkening), ii) is our census of Chemically Peculiar stars complete up to the magnitude limits we adopted ? If not, what are the  physical properties of the newly found CP stars?. 
   The abundance results for the A0-A1V sample have been published in \cite{Royer}.
   The selection criteria were: a declination higher than $-15$\degr, spectral class A0 or A1 and luminosity class V and IV and magnitudes $V$ brighter than 6.65 and a \vsini less than 65 \kms. The B8-9 sample employs the same criteria, except for the V magnitude brighter than 7.85 as these B stars are intrinsically brighter in the V band where SOPHIE reaches its maximum efficiency. Most of the stars of that B8-9 sample (40 stars) have been observed in December 2014 with SOPHIE, the \'echelle high-resolution spectrograph
   at Observatoire de Haute Provence yielding spectra coving the 3900 \AA-6800 \AA\ spectral ranger over 39 orders at a resolving power R = 75000.
   A careful abundance analysis of the high resolution high signal-to-noise ratio spectra of the A stars sample has allowed to sort out the sample of 47 A stars into 17 chemically normal stars (ie. whose abundances do not depart by more  than $\pm 0.20$ dex from solar values), 12 spectroscopic binaries and 13 chemically peculiar stars (CPs) among which five are new   CP stars. The status of these new CP stars still needs to be fully specified by spectropolarimetric observations to address their magnetic nature or by exploring new spectral ranges which we had not explored in this first study. Indeed, the abundance analysis of the A stars sample in \cite{Royer} relied only on four spectral regions: 4150--4300\,\AA, 4400--4790\,\AA, 4920--5850\,\AA, and 6000--6275\,\AA, avoiding Balmer lines and atmospheric telluric lines.
   \\
   We have now started to examine the B9-B8V sample using the full wavelength coverage provided by SOPHIE to search for new Chemically Peculiar stars. We have already reported on the discovery of 4 new HgMn stars \citep{Monier} whose spectra display strong 
 \ion{Hg}{2} lines at 3984\,\AA\ and strong \ion{Mn}{2} lines. These new HgMn stars are  HD 18104, HD 30085, HD 32867, HD 53588. In the process of our analysis of the B9-B8V sample, we have just found that HD 67044, currently classified B8 in SIMBAD, is actually another new CP star, most likely another new HgMn star and that HD 42035, classified B9V,  actually is a new spectroscopic binary containing a slow rotator which we will refer to as HD 42035 S (S for "sharp" lines) and a fast rotator, HD 42035 B (B for "broad" lines).
 A bibliographic query of the CDS for HD 67044 and HD 42035 actually reveals only 4 and 26 publications respectively.  Neither HD 67044 nor HD 42035 appear in \citeauthor{Cowley1972}'s classification (\citeyear{Cowley1972}) of the bright B8 stars.
The purpose of this paper is to report on the detection of strong   \ion{Ti}{2},  \ion{Cr}{2}, \ion{Mn}{2}, \ion{Sr}{2}, \ion{Y}{2},
\ion{Zr}{2}, \ion{Ba}{2} lines  and a weak \ion{Hg}{2} line in the spectrum of HD 67044 and perform a line identification of all the very sharp lines of HD 42035 S. We also have determined the abundances of 17 chemical elements for HD 67044 using spectrum synthesis to quantify the enhancements and depletions of these elements. As several of the lines of HD 42035 S are blended with those of HD 42035 B, we have only been able to derive provisional abundances for HD 42035 S using lines little affected by the fast rotator.

\section{Observations and reduction}
\label{sec:obs-reduc}

HD 67044 and HD 42035 have been observed at Observatoire de Haute Provence using the high resolution ($R = 75000$) mode of the SOPHIE \'echelle spectrograph \citep{Perruchot} in December 2014. 
  The  $\frac{S}{N}$ ratio of the spectra are about 130 and 174 at 5500 \AA\ respectively. The observations log is displayed in Table\,\ref{table:1}.
 The data are automatically reduced to produce 1D extracted and wavelength calibrated \'echelle orders. Each reduced order was normalised separately using a Chebychev polynomial fit with sigma clipping, rejecting points above or below 1 $\sigma$ of the local continuum. Normalized orders were merged together, corrected  by the blaze function and resampled into a constant wavelength step of about 0.02\,\AA\ \cite[see][for more details]{Royer}.
 
The radial velocities of HD 67044 and HD 42035 were derived from cross-correlation techniques, avoiding the Balmer lines and the atmospheric telluric lines. The normalized spectrum was cross-correlated with a synthetic template extracted from the POLLUX database\footnote{\url{http://pollux.graal.univ-montp2.fr}}  \citep{Palacios} corresponding to the parameters $T_\mathrm{eff}=11000$\,K, $\log g=4$ and solar abundances. A parabolic fit of the upper part of the resulting cross-correlation function yields the Doppler shift, which is then used to shift spectra to rest wavelengths. The radial velocities of HD 67044 and HD 42035 are collected in Table 2.

\section{The line spectrum of HD 67044}
\label{sec:linespec}

\begin{figure}[h!]
\vskip 0.5cm
   \centering
      \includegraphics[scale=0.33]{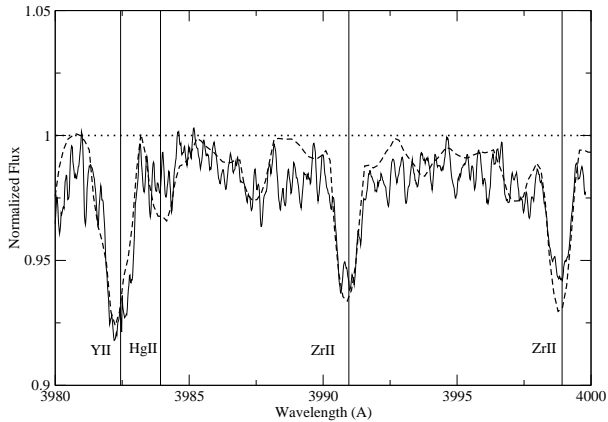}
   \caption{Comparison of the observed \ion{Y}{2} line at 3991.13 \,\AA\ , \ion{Hg}{2}  at 3983.93 \AA\ and \ion{Zr}{2} line at 3998.96 \,\AA\ of HD 67044 to a synthetic profile computed for overabundances of 1125 , 200 and 5000 times solar for yttrium, zirconium and mercury respectively.}
   \label{HD67044ZrY}
 \end{figure}

 Several spectral regions have been used to establish the chemical peculiarity of HD 67044. First, the red wing of H${\epsilon}$, which lies in order 3, harbors the \ion{Hg}{2} $\lambda$ 3983.93\,\AA\ line and several \ion{Zr}{2} and \ion{Y}{2} lines likely to be strengthened in CP stars. After proper correction for the stellar radial velocity, we found that HD 67044
does  show the \ion{Hg}{2} 3983.93\,\AA\ line as a feature absorbing about 2\% , and also strong lines of \ion{Y}{2} at 3982.59 \AA\ and of \ion{Zr}{2} at 3991.13 \AA\ and 3998.97 \AA.
The \ion{Y}{2},  \ion{Zr}{2} and \ion{Hg}{2} lines in the spectral range 3980 to 4000 \AA\ are displayed in figure \,\ref{HD67044ZrY} together with the synthetic spectrum fiting best this spectral interval.
Several other lines of \ion{Y}{2} and \ion{Zr}{2} are strong absorbers in the spectrum of HD 67044 and have been used to derive the abundances in these elements.
Second, we examined the region from 4125\,\AA\ to 4145\,\AA\  (order 6) for the \ion{Si}{2} lines at 4128.054 \AA\ and 4130.894 \AA\
and  the  \ion{Mn}{2} line at 4136.92\,\AA. In an HgMn star, the \ion{Mn}{2} line at 4136.92\,\AA\ should be strong whereas it should be absent in any comparison normal late B-type star.
Furthermore, the  lines of \ion{Mn}{2} at 4206.37\,\AA\ and 4252.96\,\AA\ should also be enhanced in the spectra of HgMn stars \citep{Gray}. These three \ion{Mn}{2} lines correspond to moderately strong  features in HD 67044, they usually are broad lines absorbing  5\% of the local continuum. Several other strong  \ion{Mn}{2} lines could be found and have been used to derive the abundance of manganese.

We find that the strongest expected lines of  \ion{Ti}{2}, \ion{Cr}{2}, \ion{Mn}{2} \ion{Sr}{2}, \ion{Y}{2}, \ion{Zr}{2}, \ion{Ba}{2} are all indeed strong features in the spectrum of HD 67044. In Table 3, we give a list of the strongest unblended lines of these species together with abundance determinations.

The presence of the \ion{Hg}{2} 3983.93\,\AA\ line and of several strong \ion{Mn}{2} and \ion{Sr}{2}, \ion{Y}{2} and \ion{Zr}{2}
lines led us to conclude that HD 67044 is actually another new HgMn star
 and should be reclassified as such. The abundance determinations presented in next paragraph do confirm this proposal.

 \begin{table}
\tiny
\caption{Observation log} 
\label{table:1} 
\centering 
\begin{tabular}{cccccc} 
\hline\hline 
Star ID & Spectral & V &Observation & Exposure & S/N \\ 
& type&  & Date & time (s ) &\\ 
\hline 
HD 67044 & B 8    & 7.48 & 2014-12-16 & 2400 & 130 \\ 
HD 42035  & B9V  & 6.55 & 2014-12-18 & 750 &  174 \\
\hline 
\end{tabular}
\end{table}


\section{The line spectrum of HD 42035}

\begin{figure}[!ht]

 \centering
   \includegraphics[scale=0.83]{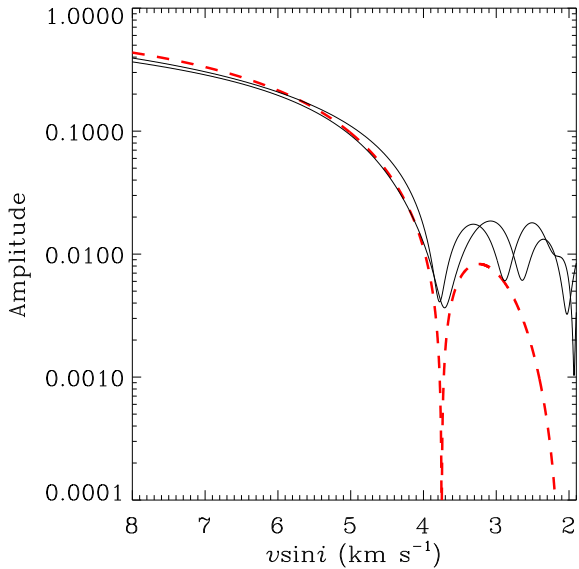}
      \caption{Fourier transforms of the \ion{Fe}{2} lines at 4508.29 and 4515.34\,\AA (solid lines) and of a synthetic profile with a $\vsini=3.7$ km\,s$^{-1}$ at a spectral resolution of 75000. On the velocity displayed on the x-axis, the position of the first zero yields the projected equatorial rotational velocity.}
         \label{FT}
   \end{figure}

\begin{figure}[th!]
\vskip 1cm
 \centering
   \includegraphics[scale=0.35]{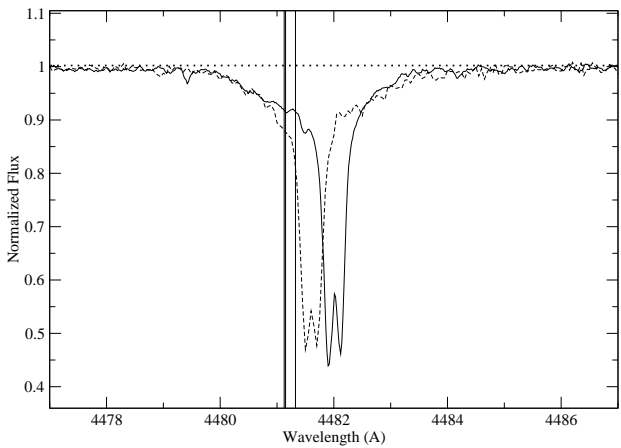}
      \caption{Comparison of  the composite \ion{Mg}{2} triplet profile  of HD 42035 taken 14 years apart: SOPHIE spectrum (solid line) and archival ELODIE spectrum (dashed line)
      .The vertical lines depict the laboratory wavelengths of the \ion{Mg}{2} triplet.}
         \label{CompHD42035}
   \end{figure}

\begin{figure}[th!]
\vskip 1cm
 \centering
   \includegraphics[scale=0.35]{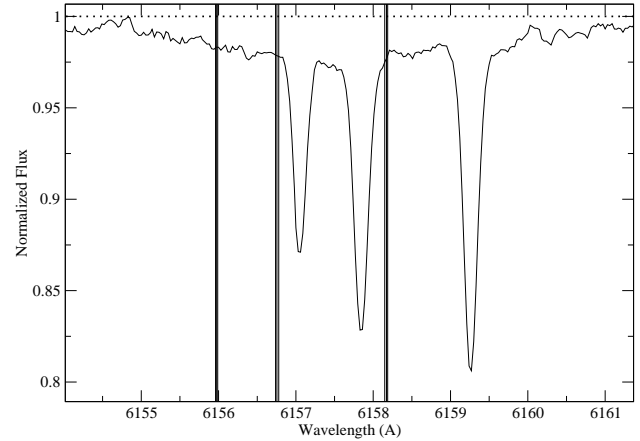}
      \caption{Composite profile of Multiplet 7 of \ion{O}{1} for HD 42035
      The vertical lines depict the laboratory wavelengths of the \ion{O}{1} lines.}
         \label{OIHD42035}
   \end{figure}

The high resolution spectrum of HD 42035 is characterised by very sharp lines originating in HD 42035 S. A first estimate of the projected equatorial rotational velocity
of HD 42035 S
has been obtained by Fourier transform analysis (Fig.\,\ref{FT}), using the two \ion{Fe}{2} lines at 4508.29 and 4515.34\,\AA. From the position of the first zero of the Fourier transform, we derived a value of $\vsini = 3.7\pm 0.2$ km\,s$^{-1}$.
A close inspection of the spectral region around the \ion{Mg}{2} triplet at 4481 \AA\ reveals that the spectrum of HD 42035 is actually a composite spectrum
containing shallow and broad lines coming from HD 42035 B onto which are overimposed redshifted very sharp lines coming from HD 42035 S. In the SOPHIE spectrum of HD 42035 obtained in December 2014, the sharp lines of the
\ion{Mg}{2} triplet are displaced, after correction for the barycentric velocity of the Earth, by about +53.0
km\,s$^{-1}$ with respect to their laboratory positions. A comparison in figure \, \ref{CompHD42035}
of our SOPHIE spectrum of HD 42035 with an archival ELODIE spectrum (R=42000) obtained in December 2000 (also corrected for barycentric velocity of the Earth) 
reveals large radial velocity variations of the sharp lines of the \ion{Mg}{2} triplet and possibly a change of the asymmetry of the broad  and shallow \ion{Mg}{2} line formed in HD 42035 B. Indeed the shallow line  had an extended blue wing in December 2014 whereas it had an extended red wing in December 2000. Recent spectroscopy obtained at DAO by E. Griffin confirms large radial velocity variations of the sharp lines and support the binary nature of this star. 
The composite nature of the spectrum is seen in many features, we show in figure \, \ref{OIHD42035} the  region of Multiplet 7 of \ion{O}{1}.


We have measured the centroids of all sharp lines of HD 42035 S absorbing more than 2 \% of the normalized flux by adjusting gaussians. This yields wavelengths accurate to 
$\pm 0.02$ \AA. 
Seven hundred and thirteen line centers were thus measured.
In order to identify these lines, a model atmosphere was computed using ATLAS9 \citep{Kurucz} with 72 layers for the effective temperature and surface gravity of HD 42035 (see par. 5.1) and a solar composition. A synthetic spectrum was then computed using SYNSPEC48 (Hubeny \& Lanz, 1992) over the entire range 3900 \AA\ to 6800 \AA\ for solar abundances
and a microturbulent velocity 0.0  km\,s$^{-1}$. It was further convolved with the ROTIN3 routine (provided with SYNSPEC48) for the FWHM of SOPHIE and the $\vsini$ of HD 42035 S.
We found that the following species are definitely present in HD 42035 S: \ion{He}{1}, \ion{C}{1}, \ion{C}{2}, \ion{Mg}{2}, \ion{N}{1}, \ion{O}{1}, 
\ion{Na}{1}, \ion{Mg}{1}, \ion{Mg}{2}, \ion{Al}{1}, \ion{Al}{2}, \ion{Si}{1}, \ion{Si}{2}, \ion{S}{2}, \ion{Ca}{1}, \ion{Ca}{2},
\ion{Sc}{1}, \ion{Sc}{2}, \ion{Ti}{2}, \ion{V}{2}, \ion{Cr}{1}, \ion{Cr}{2}, \ion{Mn}{2}, \ion{Fe}{1}, \ion{Fe}{2},
\ion{Co}{2}, \ion{Sr}{2} and \ion{Ba}{2} (see Table 5).
In contrast, the following elements have very few lines identified and therefore their presence is not confirmed:   \ion{P}{2} (only one line)
and  \ion{Nd}{2} (only one line).
We note the absence of  \ion{Y}{2} lines and that of lines of  \ion{Hg}{2} and  \ion{Pt}{2} characteristic of HgMn stars.
Also the lines of \ion{Zr}{2} are weak and we could only use one unblended line, $\lambda$ 4443.008 \AA\, to derive the zirconium abundance.
A few sharp lines have triangular profiles, they all can be identified as \ion{Fe}{2} lines\\
In table 5, we have indicated as ``broad'' and sometimes ``asymmetric'' the lines which show a broad shallow component originating in HD 42035 B which can sometimes show an assymetry.
These broad and shallow lines are due to \ion{N}{1}, \ion{O}{1}, \ion{Mg}{2},  \ion{Si}{2},  \ion{Ti}{2} and  \ion{Fe}{2}.
From these identifications, we can infer that HD 42035 B probably has an effective temperature which cannot differ much from that of HD 42035 S.A rough estimate of its projected rotational equatorial velocity has been obtained by adjusting the broad component of the composite \ion{Mg}{2} triplet profile at 4481 \AA\ with a model of
effective temperature 10740 K, $\log g$ =3.80, solar metallicity and yields $\vsini$ about 150 $\pm$ 25 km\,s$^{-1}$.

 \section{Abundance determinations}
 
 \subsection{Fundamental parameters determinations}
 
For HD 67044, we have adopted the effective temperature $T_{\rm{eff}}$ = 10519 K and surface gravity $\log g$ = 3.72 derived by \cite{Huang} from fitting the H${\gamma}$ profiles to prediction of model atmospheres. Indeed HD 67044 does not have Str\"{o}mgren's photometry which precludes applying Napiwotzky's (\citeyear{Napiwotzki}) UVBYBETA procedure to derive its fundamental parameters.
 A spectrum synthesis of the H${\gamma}$ profile was run to confirm these parameters. For HD 42035, the effective temperature, $T_{\rm{eff}}$ = 10740 K, and surface gravity, $\log g$ = 3.80,  were also taken from \cite{Huang}. Applying Napiwotzky's procedure to the Str\"{o}mgren's photometry of HD 42035 yields very similar fundamental parameters. The effective temperature obtained in this manner is probably a complex mean of the individual effective temperatures of
 HD 42035 S and HD 42035 B as the Str\"{o}mgren's  photometry most likely measured the combination of the lights of both stars.
 The adopted effective temperatures, surface gravities, projected equatorial velocities and radial velocities of HD 67044 and HD 42035 S are collected in Table\,\ref{table:2}.

 \begin{table}
\caption{Fundamental parameters}             
\label{table:2}      
\centering                          
\begin{tabular}{c c c c c}        
\hline\hline                 
Star ID & $T_\mathrm{eff}$ & $\log g$  & $v\sin i$ & $V_{rad}$\\    
            &               &               &  (km\,s$^{-1}$) & (km\,s$^{-1}$)\\
\hline                        
   HD 67044 & 10519 & 3.72 & 45.0 & 14.50 \\      
   HD 42035 S & 10740 & 3.80 & 3.7  & 53.00 \\
\hline                                   
\end{tabular}
\end{table}

 \subsection{Model atmospheres and spectrum synthesis calculations}
 
 Plane parallel model atmospheres assuming radiative equilibrium and hydrostatic equilibrium were computed using the ATLAS9 code \citep{Kurucz}. The linelist was built from \cite{Kurucz} gfhyperall.dat and its revision gfall18sep15.dat  \footnote{\url{http://kurucz.harvard.edu/linelists/gfnew/gfall28sep15.dat}} which includes hyperfine splitting levels. A grid of synthetic spectra was computed with SYNSPEC48 \citep{Hubeny} to model the
 lines of \ion{He}{1}, \ion{C}{2}, \ion{O}{1}, \ion{Mg}{2}, \ion{Al}{1} and \ion{Al}{2}, 
  \ion{Si}{2}, \ion{Ca}{2}, \ion{Sc}{2}, \ion{Ti}{2}, \ion{Cr}{2}, \ion{Mn}{2}, \ion{Fe}{2}, \ion{Sr}{2},  \ion{Y}{2}, \ion{Zr}{2}, \ion{Ba}{2}  and \ion{Hg}{2} lines. Computations were iterated varying the unknown abundance until minimization of the chi-square between the observed and synthetic spectrum was achieved. The microturbulent velocity was first assumed to be  1.5 km\,s$^{-1}$ (in agreement with the run of $v_{micr}$ with $T_\mathrm{eff}$ we established in \cite{2014psce.conf..193G}). The synthesis of the \ion{Fe}{2} lines of various strengths however imposed a lower $v_{micr} = 0$ km\,s$^{-1}$ for HD 67044  typical of an HgMn star. A null microturbulent velocity was also finally adopted for HD 42035 S after unsuccessful attempts to model the \ion{Fe}{2} lines with higher velocities.
 

\subsection{The derived abundances of HD 67044}

We have used only unblended lines to derive the abundances. There are actually few unblended lines as the projected equatorial velocity of HD 67044 broadens significantly the lines.
For a given element, the final abundance is a weighted mean of the abundances derived for each transition. These final abundances and their estimated uncertainties for HD 67044 are collected in Table 3 \cite[the determination of the uncertainties is discussed in][]{Royer}.
Table 3 contains for each analysed species the adopted laboratory wavelength, logarithm of
oscillator strength, its source, the logarithm of the absolute abundance normalised to that of hydrogen (on a scale where $\log(N_{H}) = 12$) for each transition, and the final abundance and estimated uncertainty. 
In this work, we adopted \cite{asplund2009} abundances for the Sun as a reference.
\\
The iron abundance, which is found to be about solar in HD 67044,  has been derived mostly by using several \ion{Fe}{2}  lines of multiplets 37, 38 and 186  in the range 4500--4600\,\AA\ whose atomic parameters are critically assessed in NIST\footnote{\url{http://www.nist.gov}} (these are C+ and D quality lines). These lines are widely spaced and the continuum is fairly easy to trace in this spectral region. Their synthesis always yields consistent iron abundances from the various transitions with very little dispersion. The iron abundance is probably the most accurately determined of the abundances derived here. The NLTE abundance correction for the $\lambda$ 4471.48 \AA\  \ion{He}{1} line is very small as shown by \cite{lemke} for early A-type stars and we have not corrected for it. 
We find that the following species are underabundant: He, C, O, Si, Ca, Sc while Mg, V, Fe and Ni  show mild enhancements (less than 5 times solar) and Ti, Cr, Mn, Sr, Y, Zr, Ba and Hg  show pronounced overabundances (larger than 5 times solar)., the largest overabundance being for Hg. The Sr-Y-Zr triad is inverted, yttrium beeing more abundant than strontium and zirconium. The general trend is that the heaviest elements are the most overabundant.

\subsection{The provisional abundances for HD 42035 S}

Assuming an effective temperature of 10740 K, $\log g$ = 3.80 and a solar composition, for HD 42035 S, we have derived
provisional  abundances for HD 42035 S which are collected in Table 4 The effective temperature  of HD 42035 S is probably different from this mean value so that the derived abundances are rough estimates only.
Furthermore, placing a continuum level in the composite spectrum of HD 42035 turned out to be difficult because of the presence of the rotationally broadened lines and of the continuum of HD 42035 B. The provisional abundances we derive here are therefore affected by systematic errors due to (at least) the possibility of an improper placement of the continuum
(too low  compared to what it actually is). As a consequence, the residual fluxes $\frac{F}{F_{c}}$ are likely to be too large, placing the continuum higher would decrease 
these residual fluxes and therefore lead to higher abundances. We therefore consider that the provisional abundances derived here are lower limits. The actual 
abundances of HD 42035 S must be larger than these lower limits. 
To minimize this effect  we chose to synthesize sharp lines which were as little as possible blended with the broad lines of HD 42035 B.
Realistic abundances will only be derived if we manage to disentangle the spectrum of HD 42035 S from that of HD 42035 B.
\\
We find that the light elements up to Z=21, C, O, Mg, Al, Si, S, Ca and Sc are all underabundant. The scandium deficiency is large (about 4 \% of the solar scandium abundance).
The elements heavier than Z=21 are overabundant except for Ti, Fe and Sr which are underabundant. The zirconium overabundance has been derived from only one unblended line and thus should be taken with caution. 
The absence of detection of the two strongest \ion{Y}{2} lines at 3982.60 \AA\ and 5662.92 \AA\ implies that yttrium must be solar or underabundant. 

\section{Conclusions}
\label{sec:conclusion}

We find that HD 67044, hitherto classified a normal B8 star,
displays underabundances of He, C, O, Si, Ca and Sc, mild overabundances of  Mg, V, Fe and Ni and pronounced enhancements of Ti, Cr, Mn, Sr, Y, Zr, Ba and Hg..
We therefore propose that HD 67044 actually is a new HgMn star and should be reclassified as such.  
Current monitoring of HD 67044 suggests that the \ion{Hg}{2} line at 3983.93 \AA\ is probably variable. One possible interpretation is that HD 67044 has one or several spots of overabundant Hg over its surface. More observations are planed to address this issue.
\\
The high resolution spectrum of HD 42035 appears to be a composite spectrum where redshifted very sharp lines originating from HD 42035 S are overimposed onto the broad shallow lines originating from HD 42035 B. 
The presence of a very slow rotator and a fairly fast rotator inside the same system suggests that the semi-axis of the orbit must be large.
A FT analysis of the sharp lines of HD 42035 S yields 
a very low projected equatorial velocity ($\vsini = 3.7$ km\,s$^{-1}$) which makes this star one of the very few late-B type slow rotators.
The identification of 713 very sharp lines reveals that most of the lines are due to elements up to Nickel. Strontium and Barium are also present.
Assuming a similar effective temperature and surface gravity for HD 42035 S and HD 42035 B, we have derived provisional abundances for  HD 42035 S. Light elements up to Z=22 appear to be underabundant in HD 42035 S whereas heavier elements are overabundant with the exception of Fe and Sr.
These preliminary abundances are lower limits of the actual abundances of HD 42035 S because we are probably placing the continuum too low.
A complete follow-up of this new spectroscopic binary will hopefully enable a proper disentangling of the spectra of HD 42035 S and HD 42035 B and help characterize fully the nature of each of these components, in particular the nature and abundance pattern of HD 42035 S.

\begin{acknowledgements}

We thank the referee, Dr. Fiorella Castelli, for her insightful comments which resulted into many improvements. 
We also thank the OHP night assistants for their helpful support during the three observing runs.
We acknowledge the use of the ELODIE archive at OHP.
\end{acknowledgements}

\bibliography{ref}
\bibliographystyle{aa}

\onecolumn
\begin{deluxetable}{lcccr}
\tabletypesize{\normalsize}   
\tablecaption{Elemental Abundances for HD67044. References are MA60 for \cite{MA60}, DR2006 for \cite{DR2006}, NS81 for \cite{NS81}, KZ91 for \cite{KZ91}, W96 for \cite{W96}, HBGV91 for \cite{HBGV91}, Si98 for \cite{Si98}, K79 for \cite{K79}, H99 for \cite{H99}, Ma01 for \cite{Ma01}, Sh61 for \cite{Sh61}, LD89 for \cite{LD89}, FMW for \cite{1988atpi.book.....F}, PTP for \cite{2002ApJS..138..247P}, SN14 for \cite{SN14}, NJ13 for \cite{NJ13}, T89 for \cite{T89}, RU98 for \cite{RU98}, PBL95 for \cite{PBL95}, NBS for \cite{1969AD......1....1M}, KO83 for \cite{1983AAfz...49...39K}, Bie11 for \cite{Bie11},L06 for \cite{L06}, MCS75 for \cite{MCS75}, BL48 for \cite{1948ZA.....25..325B}, Bal81 for \cite{1981ApJ...248..867B}, KL01 for \cite{KL01}, DSVD92 for \cite{DSVD92}, SR01 for \cite{SR01}, Do03 for \cite{Do03}, NIST for the online database (http://physics.nist.gov/PhysRefData/ASD/), and Kurucz (http://kurucz.harvard.edu/linelists/gfhyperall/) for the gfhyperall.dat linelist.}
\tablecolumns{5}\tablewidth{0pc}  
\tablehead{Element  & $\lambda$ (\AA) & $\log gf$ & Ref. & $\log(X/H)_{\star}$   \\}
\startdata
\ion{He}{1} & 4471.470    &-2.211  &   Ma60/DR2006      	& -1.17   \\
\ion{He}{1} & 4471.474    &-0.287 &   Ma60/DR2006     & -1.17   \\
\ion{He}{1} & 4471.474    & -1.035 &  Ma60/DR2006      & -1.17  \\
\ion{He}{1} & 4471.485    & -1.035 & Ma60/DR2006	& -1.17   \\
\ion{He}{1} & 4471.489    & -0.558 & Ma60/DR2006	& -1.17    \\
\ion{He}{1} & 4471.683    & -0.910  & Ma60/DR2006	& -1.17 \\
\ion{He}{1} & 5875.598    & -1.516  & Ma60/DR2006      &  -1.13 \\
\ion{He}{1} & 5875.613    & -0.339  & Ma60/DR2006	&  -1.13 \\
\enddata
\end{deluxetable}

\onecolumn
\begin{deluxetable}{lcccr}
\tabletypesize{\normalsize}  
\tablecaption{Provisional Elemental Abundances for HD42035 S}
\tablecolumns{5}\tablewidth{0pc}  
\tablehead{Element  & $\lambda$ (\AA) & $\log gf$ & Ref. & $\log(X/H)_{\star}$   \\}
\startdata
\ion{C}{2} &  4267.001  &  0.563 & NS81&    -3.63 \\
\ion{C}{2} & 4267.261   &  0.716 & NS81	&   -3.63 \\
\ion{C}{2} & 4267.261   &  -0.584 & NIST	&   -3.63  \\
\hline
\multicolumn{5}{c}{$[\dfrac{C}{H}]= -0.15  \pm 0.15 $ dex}\\
\hline
\ion{O}{1} & 6155.961 &  -1.363 & HBGV91	& -3.47\\
\ion{O}{1} & 6155.971 & -1.011 & HBGV91	&  -3.47\\
\ion{O}{1} & 6155.989 & -1.120 & HBGV91	&  -3.47   \\
\ion{O}{1} & 6156.737 &  -1.487 & HBGV91	&  -3.47\\
\enddata
\end{deluxetable}

\onecolumn
\begin{deluxetable}{lccccl}

\tablecaption{Proposed line identifications for HD 42035 S}
\tablecolumns{6}\tablewidth{0pc}  
\tablehead{Corr. &  Lab.  & Ident. & $E_{low}$  & $\log (gf)$ & Rem. \\ 
 $\lambda$ (\AA)  &  $\lambda$ (\AA) &  & (cm$^{-1}$) &  &   \\}
\startdata
3900.54     & 3900.550    & Ti II    &  9118.260  & -0.45 & \\
3902.95     & 3902.945  & Fe I    &  12560.933  & -0.47 &  \\
3903.26     & 3903.268  & V II     &  11908.270  & -0.89  &   \\
3903.74     & 3903.756    & Fe II   &  60807.228  & -1.49  &   \\
3905.52     & 3905.525  & Si I     &  15394.369  & -1.09   &    \\
3905.64     & 3905.644  &  Cr II   &  42986.619  &  -0.90   &    \\
3906.03     & 3906.035  & Fe II    &  44929.549  &  -1.83  &    \\
3913.46     & 3913.468  & Ti II     &  8997.710   & -0.53    &    \\
\enddata
\end{deluxetable}

\end{document}